\documentclass{PoS}
\usepackage{wrapfig}

\title{Entanglement entropy in SU(N) gauge theory }

\ShortTitle{Entanglement entropy in SU(N) gauge theory }

\author{\speaker{Alexander Velytsky}\\
        Enrico Fermi Institute, University of Chicago, 5640 S. Ellis Ave., Chicago, IL 60637, USA, \\
        HEP Division and Physics Division, Argonne National Laboratory, 9700 Cass Ave., Argonne, IL 60439, USA\\
        E-mail: \email{vel@theory.uchicago.edu}}
\abstract{The entanglement entropy of $SU(N)$ lattice gauge theory is studied exactly in $1+1$ space-time dimensions and in Migdal-Kadanoff approximation in higher dimensional space. The existence of a non-analytical behavior reminiscent of a phase transition for a characteristic size of the entangled region is demonstrated for higher dimensional theories.
}

\FullConference{The XXVI International Symposium on Lattice Field Theory\\
		 July 14-19 2008\\
		 Williamsburg, Virginia, USA}

\begin{document}

\section{Introduction}
Given a quantum state $|\psi\rangle$ and a corresponding density matrix $\rho=|\psi\rangle\langle\psi|$ let us introduce two observers $A$ and $B\equiv \bar A$, such that observer $B$ sees only degrees of freedom complimentary to degrees of freedom seen by observer $A$.
Then one can form a reduced density matrix for observer $A$: $\rho_A={\rm Tr}_B \rho$,
and define the entanglement entropy as the von Neumann entropy of reduced states
\begin{eqnarray}
S_A=-{\rm Tr}_A\rho_A \log \rho_A=-\sum_i\lambda_i\log\lambda_i,
\end{eqnarray}
where $\lambda_i$ are eigenvalues of the reduced density matrix. 
The entanglement entropy $S_A=0$ for a product state, and is at maximum for a maximally entangled state: $0\le S_A \le 1 [ebit]$.

As a simple example consider a bipartite system
$|\Psi\rangle=\cos\theta|\uparrow_A\rangle\otimes|\downarrow_B\rangle+\sin\theta|\downarrow_A
\rangle\otimes|\uparrow_B\rangle$.
The reduced density matrix is
$\rho_A=\cos^2\theta|\uparrow\rangle\langle\uparrow|+\sin^2\theta|\downarrow\rangle
\langle\downarrow|$,
while the entanglement entropy is
\begin{equation}
S_A=-2\cos^2\theta\log\cos\theta-2\sin^2\theta\log\sin\theta,
\end{equation}
and takes its maximum value of $\log 2$ when $\cos^2\theta=\frac12$.

\begin{wrapfigure}{r}{0.34\columnwidth}
\includegraphics[width=0.33\columnwidth,height=4cm]{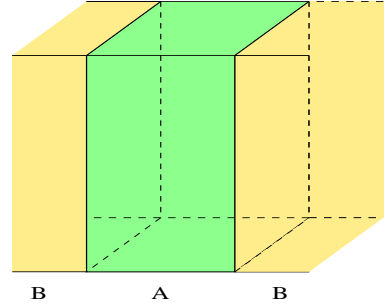}
\caption{Partitioning by two imaginary surfaces into regions $A$ and $B$.}
\label{fig:geom}
\end{wrapfigure}
One of the more interesting application of the entanglement entropy is its use as a probe of phase, which is especially important in situations where an order parameter is not known.
It turns out that in confining gauge theories the entanglement entropy may show an interesting non-trivial behavior. First it was studied in gravity duals of confining large $N$ gauge theories 
\cite{Klebanov:2007ws}, where the following geometry was considered (see Fig. \ref{fig:geom})
\begin{eqnarray}
A&=&\mathbb{R}^{d-1}\times \mathbb{I}_l,\nonumber\\
B&=&\mathbb{R}^{d-1}\times (\mathbb{R}- \mathbb{I}_l),
\label{eq:region}
\end{eqnarray}
here $\mathbb{I}_l$ is a line segment of length $l$. It was observed that at length separation $l=l_c$
the entanglement entropy exhibits a non-analytical change in behavior,  reminiscent of a phase transition. 

\begin{wrapfigure}{r}{0.33\columnwidth}
\includegraphics[width=0.33\columnwidth,height=5cm]{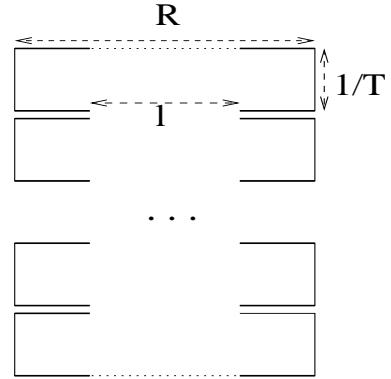}
\caption{$Z_n$ for $1+1$ dimensional gauge theory.}
\label{fig:zn2d}
\end{wrapfigure}
A natural question to ask if this behavior is relevant for small $N$ gauge theories. This question was first addressed in Migdal-Kadanoff formalism \cite{Velytsky:2008rs}. Subsequently other measures of entanglement (purity) for $SU(2)$ gauge theory were studied in Monte Carlo simulations 
\cite{Buividovich:2008kq}. The results of the Migdal-Kadanoof study are reviewed in this proceedings.

The expression for the von Neumann entanglement entropy is obtained using the replica trick 
\cite{Callan:1994py,Calabrese:2004eu}: $n$ replicas of the system 
(with region $B$ integrated out) are glued along the time boundaries of region $A$. 
We demonstrate the procedure for a $2d$ system in Fig. \ref{fig:zn2d}. Taking the trace of the combined system we obtain its normalized partition function 
\begin{equation}
{\rm Tr}\rho_A^n=\frac{Z_n(A)}{Z^n},
\end{equation}
where $Z=Z_1$ is the partition function of the original system. Then after performing analytical continuation to real $n$ one obtains the entanglement entropy 
\begin{equation}
S_A=-\lim_{n\rightarrow 1}\frac\partial{\partial n}{\rm Tr}\rho_A^n
=-\lim_{n\rightarrow 1}\frac\partial{\partial n}\frac{Z_n(A)}{Z^n}.
\end{equation}

We apply the described technique to $SU(N)$ gauge theory in $D=d+1$ dimensions
\begin{equation}
Z=\int\prod_ldU_l \prod_p e^{-S_p},
\label{eq:pf}
\end{equation}
where $S_p\equiv S(U_p)=-\beta/(2N){\rm Tr}U_p+h.c.$ is a standard Wilson lattice action with
the plaquette variable $U_p=\prod_{l\in\partial p}U_l$ and 
$\beta=2N/g^2$ is the inverse lattice coupling.
The gauge invariant action is a class function and therefore allows for character expansion
\begin{equation}
e^{-S_p}=\sum_r F_r d_r\chi_r(U_p)\equiv F_0\left(1+\sum_{r\neq0}c_rd_r\chi_r(U_p)\right),
\label{eq:char}
\end{equation}
where $c_r=F_r/F_0<1$ and $F_r=\int dU e^{-S(U)}\frac{1}{d_r}\chi^*_r(U)$.

\section{$SU(N)$ gauge theory in $1+1$ dimensions: Exact solution}
We start with exactly solvable 2-dimensional $SU(N)$ gauge theory (for an overview and large $N$ treatment of zero temperature $U(N)$ gauge theory see Ref. \cite{Gross:1980he}). At finite 
temperature the gauge theory leaves on an $\mathbb{R}\times \mathbb{S}_1$ manifold compactified in time direction with period $1/T$. The corresponding lattice theory is formulated on an $N_r\times N_t$ 
time direction periodic lattice, with space-time cut-off $a$ and $a N_t=1/T$ and $aN_r=R$.

After the partition function is character expanded (\ref{eq:char}) one can consider various contributions to it from minimal surface elements bounded by a single loop $\partial A$
\begin{equation}
f(\{a\}; \partial A)\equiv 1+\sum_{i\neq0}a_id_i\chi_i(\partial A),
\end{equation}
The contribution to the partition function of two such surface elements $A$ and $B$ with a common boundary $A\cap B$, which is integrated out,  defines a new elementary surface
\begin{center}
\includegraphics[width=0.4\columnwidth,height=0.7cm]{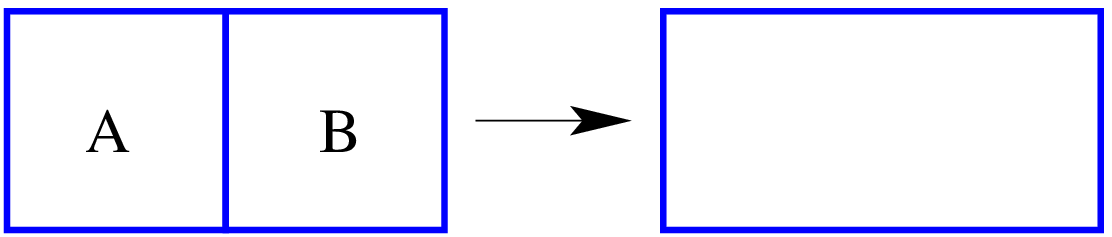}
\end{center}
\begin{equation}
f(\{c\}; \partial (A\cup B)) = \int d(A\cap B) f(\{a\}; \partial A) f(\{b\}; \partial B) 
=  1+\sum_{i\neq0}c_id_i\chi_i(\partial (A\cup B)),
\label{eq:join}
\end{equation}
where $c_i=a_ib_i$.
Thus the junction of the surfaces in the space of character coefficients is represented by an ordinary product.

Now for any 2-dimensional surface one can integrate all the internal links (in this way joining elementary surfaces). The resulting expression for the partition function is
\begin{equation}
Z=\int\prod_{l\in \partial A}dU_l  \sum_r F_r^A d_r\chi_r(U_{\partial A}),
\label{eq:pf2}
\end{equation}
where $A=N_r N_t$ is the area of the total surface in plaquette units and $\partial A$ is the contour enclosing the surface. A similar result follows for the partition function of the glued system 
$Z_n$, but with the corresponding surface area $A_n=n A=n N_r N_t$ and perimeter $\partial A_n$.

In order to evaluate the perimeter integral one needs to set the spatial boundary conditions (BC). 
The free BC produces a trivial result $Z= F_0^A$ and $S_A= 0$. Therefore we set the periodic BC. The perimeter integrals for $Z$ and $Z_n$ result in
\begin{eqnarray}
Z&:&\int dV\int dU \chi_r(UVU^\dagger V^\dagger)=\int dV \frac 1{d_r}\chi_r(V)\chi_r(V^\dagger)=\frac1{d_r},\\
Z_n&:&\int dU_1...dU_n\frac1d_r\frac{\chi_r(U_1)...\chi_r(U_n)}{d_r^{n-1}}\frac{\chi_r(U_1^\dagger)...\chi_r(U_n\dagger)}{d_r^{n-1}}=\frac1{d_r^{2n-1}},
\end{eqnarray}
and one can obtain the entanglement entropy 
\begin{equation}
S_A=-\left.\frac\partial{\partial n}\frac{Z_n}{Z^n}\right|_{n=1}
=\log(1+\sum_{r\neq 0}c_r^A) -\frac{\sum_{r\neq0}c_r^A\log c_r^A/d_r^2}{1+\sum_{r\neq 0}c_r^A}.
\label{eq:sa}
\end{equation}
It is interesting to note that $S_A(l)$ is $l$-independent $| l\neq0$, and its behavior is similar to end-point phase transition $S_A(l=0)=0$.

\begin{wrapfigure}{r}{0.5\columnwidth}
\includegraphics[width=0.5\columnwidth]{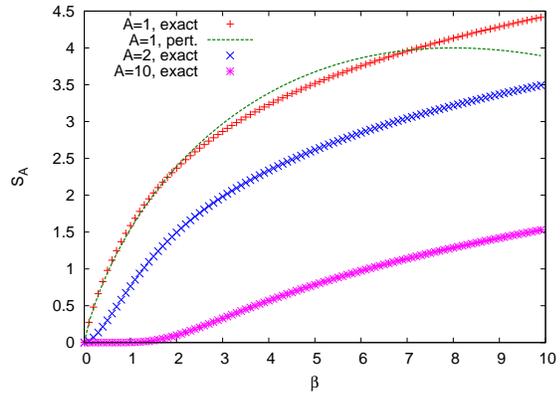}
\caption{$SU(2)$ entanglement entropy at various lattice $\beta$ and surface area $A$.}
\label{fig:su2_ee}
\end{wrapfigure}
In the expression for the entanglement entropy (\ref{eq:sa}) one can truncate the series 
if $A>>1$ or, equivalently, if $\beta$ is small. 
Let us be more specific and consider $SU(2)$ gauge theory. Then $c_r=I_{2r+1}(2\beta)/I_1(2\beta)$, where $I_n(x)$ is the modified Bessel function. In strong coupling limit
\begin{eqnarray}
I_{2r+1}(2\beta)\approx\frac1{\Gamma(2r+2)}\beta^{2r+1}, \\\nonumber
{\rm where}\quad 0<2\beta<<\sqrt{2r+2},
\end{eqnarray}
and $c_r=\beta^{2r}/(2r+1)!$.
One can then sum terms of expression (\ref{eq:sa}) to a given precission to get
 \begin{equation}
 S_A=\epsilon(1+\log 4-\log\epsilon), \quad \epsilon=\left(\frac\beta 2\right)^A.
\end{equation}

We present the results for the entanglement entropy of $1+1$ dimensional $SU(2)$ gauge theory as a function of the lattice coupling for various surface areas $A=1,2$ and 10 in Fig. \ref{fig:su2_ee}. For $A=1$ we also plot the strong coupling perturbation result.

\pagebreak

\subsection{Higher dimensions: Migdal-Kadanoff treatment}
\begin{wrapfigure}{r}{0.3\columnwidth}
\includegraphics[width=0.3\columnwidth]{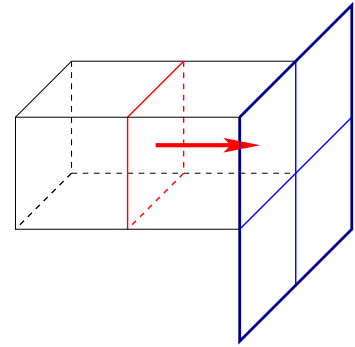}
\caption{\label{fig:mk} Illustration of the Migdal-Kadanoff procedure.}
\end{wrapfigure}
The higher dimensional $SU(N)$ gauge theories cannot be solved exactly, however the Migdal-Kadanoff (MK) approximate approach is known to produce robust results in studies of phase structure of the models.  
The standard MK decimation procedure ($\lambda$-transformation) in $D=d+1$ dimensions is defined
by the following recursive steps (see Fig. \ref{fig:mk})
\begin{eqnarray}
\label{eq:MK}
e^{-S'_p(U)}&=&\left[\sum_r F_r^Ad_r\chi_r(U)\right]^{\zeta^{1-b}}, \\\nonumber
\quad F_r&=&\int dU e^{-\zeta^{b}S_p(U)}\frac{1}{d_r}\chi^*_r(U).
\end{eqnarray}
Here $\lambda$ is the scaling factor of the renormalization group (RG) transformation,
$\zeta=\lambda^{D-2}$ is the factor by which we strengthen the interaction on the resulting coarser lattice and
$A=\lambda^2$ is the surface of the new elementary plaquette in units of plaquettes of the underlying fine lattice. One can choose between two orderings of bond moving and strengthening of the couplings steps: $b=0$ corresponds to Migdal, while $b=1$ to Kadanoff prescription.

As an example consider a gauge theory formulated in a $2+1$ dimensional box. One can then  
decimate out all bulk degrees of freedom and be left only with the perimeter integral. 
Setting free boundary conditions one then gets
\begin{eqnarray}
Z&=&\int dU dV f(\{c_{xy,i}\};U^\dagger V U V^\dagger) f(\{c_{t,i}\};V)
=1+\sum_{i\neq0}c_{xy,i}+\sum_{i,j\neq0}c_{xy,i}d_j c_{t,j}D^i_{ij},
\label{eq:zcube}
\end{eqnarray}
where 
\begin{equation}
f(\{c_{z}\}; \partial A)\equiv 1+\sum_{i\neq0}d_ic_{z;i}\chi_i(\partial A_z),\quad z=\pm x,\pm y,t
\end{equation}
and
\begin{equation}
D^k_{ij}=\int dV \chi_k(V^\dagger)\chi_i(V)\chi_j(V)=
\left(
\begin{array}{c}
  k   \\
  n_1  \mu
\end{array}
\right)
\left(
\begin{array}{c}
  k   \\
  \nu n_1
\end{array}
\right)^*
\left(
\begin{array}{ccc}
  k& i  &j   \\
  \mu& n_2  & n_3  
\end{array}
\right)^*
\left(
\begin{array}{ccc}
  k& i  &j   \\
  \nu& n_2  & n_3  
\end{array}
\right)
\end{equation}
are coefficients of the Clebsch-Gordan series ${\mathcal D}^{(i)}\times {\mathcal D}^{(j)}=\sum_k D^k_{ij} {\mathcal 
D}^{(k)}$ for the Kronecker product of irreducible representations (see \cite{Velytsky:2008rs} for details).

\begin{wrapfigure}{r}{0.39\columnwidth}
\centering
\includegraphics[width=0.33\columnwidth,height=3.6cm]{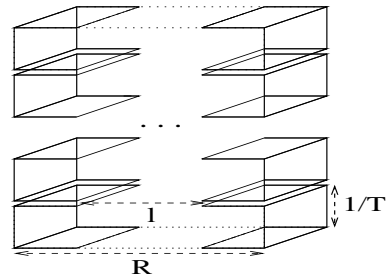}
\caption{$Z_n$ for $2+1$ dimensional theory.}
\label{fig:zn3d}
\end{wrapfigure}
Now we can apply MK procedure in order to compute the partition functions for  ordinary $Z$
and glued $Z_n$ systems, see Fig. \ref{fig:zn3d}. In order to cancel out the contribution from the bulk 
we cary out decimations for $Z_n$ and $Z$ in exactly the same manner. First we start with the standard MK decimation procedure ($\lambda$-transformation) (\ref{eq:MK}).
The decimation should be altered
when the lattice spacing becomes equal to $l$ (the smallest scale in the problem).  This affects only $l$-plaquettes, which have one link spanning the entangled region. In this case one still can move plaquettes in $D-2$ direction but the tiling is done with $\lambda$ plaquettes
\begin{equation}
\label{eq:MKl}
e^{-S'_{p;l}(U)}=\left[\sum_r F_r^\lambda d_r\chi_r(U)\right]^{\zeta^{1-b}},\quad
\quad F_r=\int dU e^{-\zeta^{b}S_{p;l}(U)}\frac{1}{d_r}\chi^*_r(U).
\end{equation}
We refer to such a procedure as $\rho$-transformation. 
All the other plaquettes are unaffected by this change and are decimated according to the standard
($\lambda$-transformation) procedure. 

If we choose $l$ to extend in $y$ direction, we can write the final formula for the entanglement entropy \cite{Velytsky:2008rs}
\begin{equation}
S_A=-\dot{\tilde F}_{t,0}+\log Z -\frac{\dot f_n}Z
\label{eq:sa_fin}
\end{equation}
where the dot stands for $\dot X=\left.\frac\partial{\partial n}X\right|_{n=1}$ and 
\begin{eqnarray}
Z&=&1+\sum_{i\neq0}(c^s_{x,i}\bar c^s_{x,i}c_{y,i})^2+\sum_{i,j\neq0}(c^s_{x,i}\bar c^s_{x,i}c_{y,i})^2d_j c^s_{t,j}\bar c^s_{t,j}D^i_{ij},\\
\tilde{F}^s_{t,j}&=&\int dU\left(1+\sum_{i\neq0}d_ic^s_{t,i}\chi_i(U)\right)^n\frac1{d_j}\chi_j(U^\dagger),\\
\dot{f}_n&=&\sum_{i\neq0}(c^s_{x,i}\bar c^s_{x,i}c_{y,i})^{2}\left(
\log\frac{(c^s_{x,i}\bar c^s_{x,i}c_{y,i})^{2}}{d_i^4}\left(1+\sum_{j\neq0}d_j \tilde 
c_{t,j}D^i_{ij}\right)
+\sum_{j\neq0}d_j \dot{\tilde c}_{t,j}D^i_{ij}\right).
\end{eqnarray}

The final expression appears to be very complicated. However, one can analyze its behavior as a function of $l$. First we simplify the expression by choosing the geometry with specific symmetry, so that $c^s_{t,i}=c^s_{x,i}=c^s_i$. Note that the dependance on $l$ is encoded in the values of $c^s_{i}$.
Note that $l$ regulates when $\lambda$-transformation is switched to $\rho$-transformation, i.e. it sets the initial value for $c^s_{i}(m_0)$ under $\rho$-transformations.
Next we analyze the RG flow of $SU(2)$ gauge theory for $c^s_{i}(m)$ as a function of number of iterations $m$ under Migdal recursion depending on the starting point.
We simplify the numerical simulation by considering a starting action in the wilsonian form on $N_{l,t}=1$ lattice. Note however that even after the first iteration step the action is of a single plaquette form, but generally has all irreducible representations.

\begin{figure}[ht]
\centering
\includegraphics[width=0.6\columnwidth]{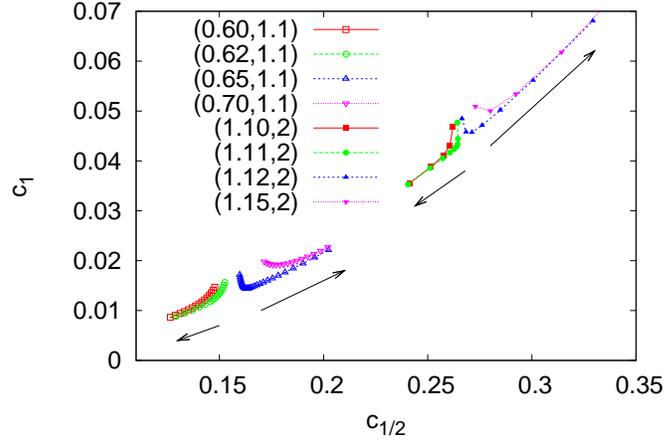}
\caption{Migdal decimation flow for $3+1$ dimensional $SU(2)$ gauge theory. Projection to  $c^s_{1/2}$ and $c^s_1$; $(\beta,\lambda)$ are indicated.}
\label{fig:flowm4d}
\end{figure}
We show the projection of the RG flow on the fundamental-adjoint plain in Fig. \ref{fig:flowm4d}. We observe that depending on the starting coupling $\beta$, which sets the $l$-scale, the flow is in different directions. This is a clear indication of a phase transition like behavior. For the scale factor we choose 
$\lambda=1.1$, which is known to map correctly the mixed action phase diagram 
\cite{Bitar:1982bp}. The transition
occurs at $\beta_c\in 0.(62,0.65)$, which corresponds to length scale $l^*_c$
\begin{equation}
l^*_c/l_c\in(1.56,1.66),
\end{equation}
where $l_c=1/T_c$ is the QCD scale.

\section{Summary}
We studied the entanglement entropy in $d+1$ $SU(N)$ gauge theory. 
 The $d=1$ theory was solved exactly. Setting periodic BC we obtained a non-zero universal value for the entanglement entropy, which is independent of the size $l$ (end-point transition).

Using MK decimation we approximately computed the ratio of partition functions and entanglement entropy for $d\ge2$. For $3+1$ $SU(2)$ gauge theory we demonstrated that there is a non-analytical change in the RG flow of character coefficients $c$ which define $S_A$. This allowed us to find the length scale where the transition occurs $l^*_c/l_c\in(1.56,1.66)$, which is comparable to the value 
$l^*_c/l_c=2$ derived for infinite $N$ theory \cite{Klebanov:2007ws}.

It is worth to point out that other measures of entanglement, such as the purity $\mu=Tr\rho^2$, 
Tsallis entropy $S_q=1/(q-1)(1-Tr\rho^q)$  and R\'enyi entropy
$H_1=1/(1-q)\log Tr\rho^q$ (specifically at $q=2$) are all dependent on $c_i^s$ and therefore will show non-analyticity in the RG flow.

\acknowledgments This work is supported in part by the U.S. Department of Energy, Division of High Energy Physics and Office of Nuclear Physics, under Contract DE-AC02-06CH11357, and in part under a Joint Theory Institute grant.

\end{document}